\documentclass[aps,prl,twocolumn,groupedaddress,nofootinbib]{revtex4}
\usepackage{graphicx,color}
\usepackage[utf8]{inputenc}
\usepackage[english]{babel}
\usepackage{amsmath,amsfonts,amssymb}
\def\phi{\varphi}

\newcommand{\vc}[1]{\boldsymbol{#1}}
\newcommand{\vnab}{\vc{\nabla}}
\newcommand{\pder}[2]{\frac{\partial #1}{\partial #2}}		% partial derivative
\newcommand{\fder}[2]{\frac{\delta #1}{\delta #2}}		% functional derivative

\renewcommand{\r}{\vc{r}}
\renewcommand{\k}{{\vc{k}}}
\newcommand{\q}{{\vc{q}}}
\newcommand{\vs}{{\vc{v}_\text{s}}}

\newcommand{\dn}{\delta n}
\newcommand{\dph}{\delta \phi}
\newcommand{\revision}[1]{{#1}}
\renewcommand{\Im}{\text{Im}}
\renewcommand{\Re}{\text{Re}}
\begin{document}
\begin{abstract}
We consider elementary excitations of an interacting
Bose-Einstein condensate in the mean-field
framework. As a building block for understanding the dynamics of systems
comprising interaction and disorder,
we study the scattering of Bogoliubov excitations by a 
single external impurity potential. A numerical
integration of the Gross-Pitaevskii equation shows that the
single-scattering amplitude has a marked angular anisotropy. By a
saddle-point expansion of the hydrodynamic mean-field energy
functional, we derive the relevant scattering amplitude including the
crossover from sound-like to particle-like excitations.  
The very different scattering properties of these limiting cases are
smoothly  connected  by an angular envelope function with a
well-defined node of vanishing scattering amplitude. We find that the
overall scattering is most efficient at the crossover from phonon-like
to particle-like  Bogoliubov excitations. 
\end{abstract}

\title{Anisotropic scattering of Bogoliubov excitations}

\author{Christopher Gaul}
\email{christopher.gaul@uni-bayreuth.de}
\author{Cord Axel M\"uller}
\affiliation{Physikalisches Institut, Universit\"at Bayreuth, D-95440 Bayreuth, Germany}

\maketitle 

\revision{Below a critical temperature Bose gases undergo a phase
transition and form a Bose-Einstein condensate (BEC)
\cite{Dalfovo1999,Hadzibabic2006,Krueger2007}.
It is a long-standing
question how such a condensate is influenced by the competition
between inter-atomic interactions on
the one hand and external disorder on the other
\cite{Giamarchi87,Fisher89}.}  
%In disordered optical potentials the BEC undergoes
%crossovers\footnote{In \cite{Lugan2007} the term crossover is used
%instead of quantum phase transition.} from superfluid to a fragmented
%BEC (Bose glass) or even a Lifshits glass
%\cite{Lugan2007,Lewenstein2007}.}  
Generically, both the ground-state phase diagram and non-equilibrium
features depend crucially on the presence of soft modes and their
properties \cite{Belitz2005,Gurarie2003}. 
At low temperatures, the relevant excitations of a BEC are Bogoliubov
excitations \cite{Posazhennikova2006} \revision{with collective
properties due to the repulsive inter-atomic interaction.} Their
dispersion relation interpolates between the collective sound-wave and
the single-particle regimes. In gaseous BEC, these Bogoliubov
excitations can be \revision{experimentally} created and analysed
using Bragg spectroscopy
\cite{Stamper-Kurn1999,Vogels2002,Steinhauer2002,Steinhauer2003}. 

\revision{In this article we focus on the controlled 2D scattering of
Bogoliubov excitations by a single elementary impurity.}
We present numerical evidence for highly anisotropic scattering that features a characteristic node. 
By a variational treatment of the quantum hydrodynamic energy
functional, we derive the effective Hamiltonian for impurity
scattering and obtain analytical expressions for the scattering
amplitude and the position of the node. 
\revision{We
expect our results to be useful in the future for understanding the 
quantum-transport dynamics in disordered media, which builds on repeated
single-scattering events \cite{Rossum1999}. 
}

\begin{figure}[bthp]
\begin{center}
\begingroup
\setlength{\unitlength}{\linewidth}%
\begin{picture}(0.46,0.46)(0,0)%
\put(0,0.06){\includegraphics[width=0.5\linewidth]{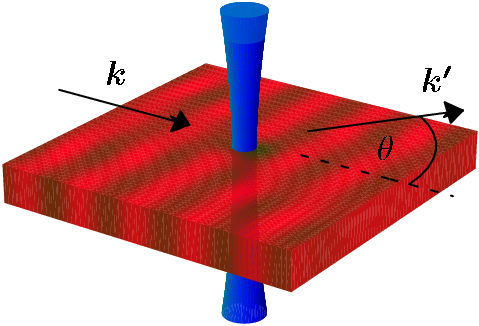}}
\put(0.01,0.02){(a)}
\end{picture}
\endgroup
\hfill
\begingroup
\setlength{\unitlength}{\linewidth}%
\begin{picture}(0.46,0.46)(0,0)%
\put(0,0){\includegraphics[width=0.46\linewidth]{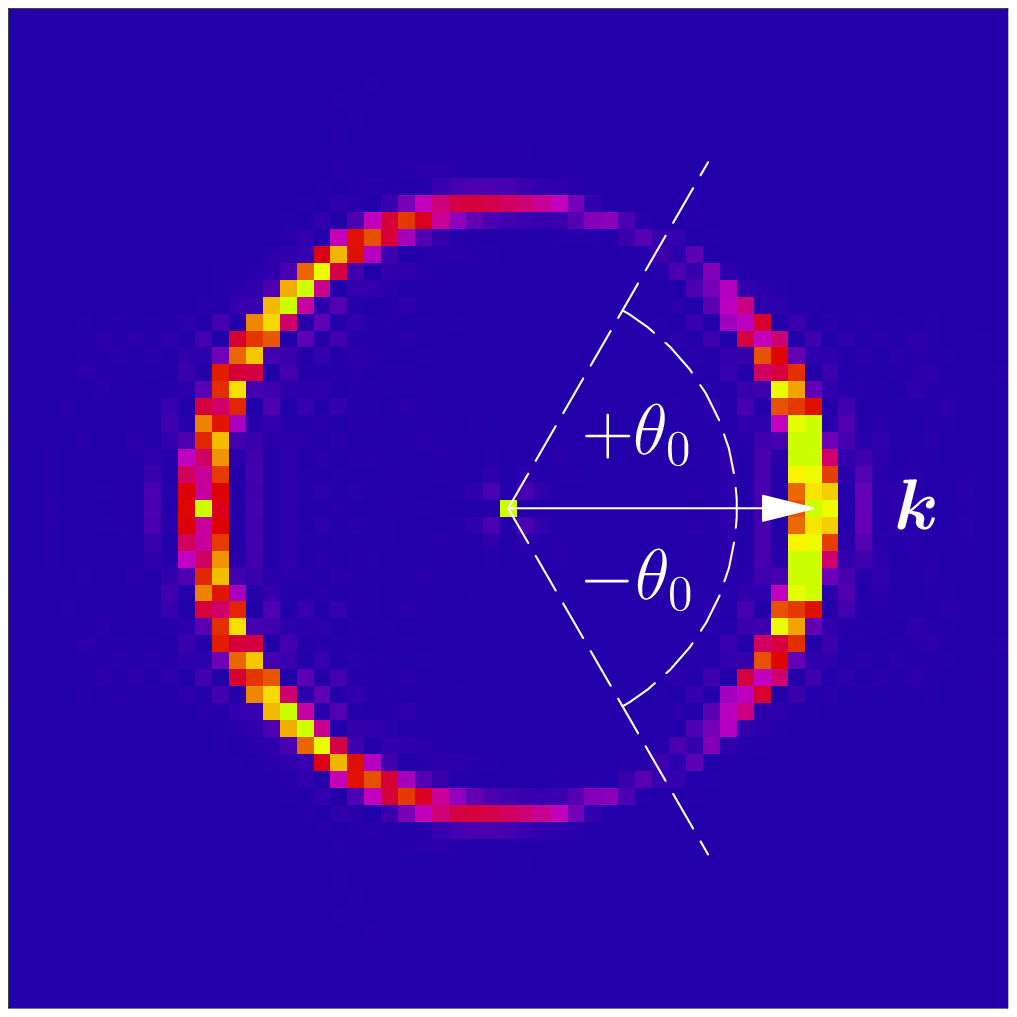}}
\put(0.01,0.02){\color{white}(b)}
\end{picture}
\endgroup
\caption{(a) A blue-detuned laser beam focused
perpendicularly through a 2D BEC provides an
impurity potential $V(\r) = \mathcal{V} \exp(-\r^2/r_0^2)$ that scatters an
incident Bogoliubov wave with wave vector $\k$ into modes $\k'$ at
scattering angle~$\theta$. (b) Density plot $|\delta \Psi_{\k'}|^2$ of
the scattered state obtained by numerical integration of the
GP equation. The components of the scattered wave are distributed on the elastic
circle $|\k'| = k$ with characteristic nodes at
$\pm\theta_0$. For the numerical parameters $\mathcal{V} = 0.25 \mu$, \mbox{$k
\xi=1$}, $k r_0 = 0.5$, scattering is suppressed around
$\theta_0\approx\pi/3$.
}\label{figSetupDensity} 
\end{center} 
\end{figure}

\section{Simulation}
We start our investigation by 
numerically simulating the scattering of Bogoliubov excitations by an
impurity potential in a two-dimensional BEC,   
see Fig.~\ref{figSetupDensity}(a), 
\revision{under periodic
boundary conditions that mimic the very shallow trap required for an
experimental realisation}.  
At temperatures much lower than the critical temperature
\cite{Dalfovo1999,Hadzibabic2006,Krueger2007}, 
a mean-field description in terms of the
macroscopically occupied wave function $\Psi(\r,t)$ is appropriate.
First, we obtain the 
\revision{static} 
ground-state density $n_0(\r)=|\Psi_0(\r)|^2$ of
the condensate
in presence of an impurity potential 
$V(\r) = \mathcal{V} \exp(-\r^2/r_0^2)$ by imaginary time propagation of the
Gross-Pitaevskii (GP) equation \cite{Pitaevskii2003,Dalfovo1999}.
Then a plane wave Bogoliubov excitation
with wave vector $\k$ is 
\revision{superimposed on this 
ground state solution by a suitable choice of initial conditions at
time $t=0$.} 
Then the ensuing time evolution according to the full GP equation
is calculated.  
During the simulation the Bogoliubov wave moves forward and is
scattered at the impurity. 

In order to analyse the scattered state, we Fourier transform a
snapshot of the deviation $\delta\Psi(\r,t)=\Psi(\r,t)-\Psi_0(\r)$
from the ground state around the impurity. 
Fig.~\ref{figSetupDensity}(b) shows the resulting momentum density $|\delta
\Psi_{\k'}|^2$. The scattering is essentially elastic, with
the components of the scattered wave distributed on the circle 
$|\k'| = k$. Surprisingly
at first sight, scattering is suppressed at two symmetric 
angles $\pm\theta_0$ with respect to the forward direction. 
In the deep sound-wave regime
$k\xi\ll 1$ (where $\xi$ is the healing length), we find these nodes at $\theta_0\to \pi/2$ 
resulting in a dipole scattering (p-wave) characteristic.
For large values of $k\xi$, when the Bogoliubov excitations are
particle-like, the nodes shift to the forward direction, $\theta_0\to
0$. 
This intriguing anisotropic scattering of Bogoliubov excitations is a signature of 
the intricate crossover from single-particle to
collective excitations in interacting Bose-Einstein condensates. 
  
\section{Limiting cases}
Before tackling the full quantum hydrodynamical problem, we find it instructive to discuss the two limiting cases of pure particle-like excitations ($k\xi\gg1$) and pure
sound-like excitations ($k\xi\ll 1$), where the expected angular distribution can be derived from elementary considerations. 

In the single-particle part of the Bogoliubov spectrum, excitations
are plane matter waves with dispersion relation $\epsilon_k^0 = \hbar^2k^2/2m$. The
amplitude of a single-scattering process $\k \mapsto \k' = \k+\q$ is proportional 
to the Fourier component 
$V_\q = L^{-d/2}\int d^d r V(\r)e^{-i\q\cdot\r}$\label{textFTconvention}
of the scattering potential. 
If the potential $V(\r)$ varies on a
characteristic length $r_0$, the scattering may be anisotropic if the
wave can resolve this structure, $qr_0
\ge 1$ 
\cite{Kuhn2005,Kuhn2007a}. In the opposite case of a small obstacle such that
$qr_0\ll1$, also known as the s-wave scattering regime, the
scattering amplitude is simply proportional to $V_0$ and can therefore only be \emph{isotropic}.

Quite on the contrary, we expect a very \emph{anisotropic} scattering
amplitude, proportional to
$V_0\cos\theta$, 
in the deep sound-wave part of the Bogoliubov
spectrum. Indeed, in the Thomas-Fermi regime, where the 
healing length $\xi$ is much smaller than the scale $r_0$ of
potential variations, the condensate ground-state density follows the
potential: 
$n_\text{TF}(\r)=\left[ \mu - V(\r)\right]/g$. 
(Here as in the following, we assume that the 
impurity  potential is always smaller than the chemical potential 
$\mu$, such that $n > 0$ everywhere.)
Excitations of the superfluid ground state in the  regime
$k\xi\ll 1$ are longitudinal sound waves with density fluctuations $\dn
= n - n_\text{TF}$ and phase fluctuations $\dph$. Importantly, the
phase is the potential for the local superfluid velocity $\vs=(\hbar/m)\vnab \dph$.  
The superfluid hydrodynamics is
determined by the continuity equation 
%\begin{equation} 
$\partial_t n + \vnab \cdot (n\vs)=0$
%\end{equation}
on the one hand, and by the Euler equation for an ideal compressible
fluid, 
%\begin{equation}
$ m \left[\partial_t\vs + (\vs\cdot\vnab)\vs\right] = -\vnab(g n +
V)$, 
%\end{equation}
on the other.
To linear order in $\dn$ and $\dph$, these two equations
can be combined
to a single wave equation 
\begin{equation}\label{eqMinimalWave}
m \left[c^2 \nabla^2 -\partial_t^2 \right] \dn = 
\vnab \cdot \bigl(V(\r) \vnab \dn\bigr),  
\end{equation}
where the sound velocity appears as $c = \sqrt{\mu/m}$. 
The excitations of a 
homogeneous fluid ($V(\r)=0$) are plane sound waves 
%$\dn(\r,t) = \dn_0 e^{i(\k\cdot\r - ckt)} + c.c.$  
with linear dispersion $\omega_k= c\, k$.  
The gradient-potential operator %$\vnab\cdot V(\r)\vnab$  
on the right-hand side %of equation \eqref{eqMinimalWave} 
then causes scattering with an amplitude proportional to 
$-(\k\cdot\k') V_{\k'-\k}$.
Hence, the potential component $V_{\k'-\k}$, which must appear in all
cases to satisfy momentum conservation, is multiplied with a dipole (or p-wave) characteristic
$A(\theta)= - \cos\theta$. 
This scattering cross-section with a node at
$\theta_0=\pm \pi/2$ can be understood, in the frame of reference where
the local fluid velocity is zero, as the dipole
radiation pattern of an 
impurity that oscillates to and fro, quite similar to the case of classical
sound waves scattered by an impenetrable obstacle 
\cite{Landau2004}. 

\revision{
By continuity, there should be a smooth crossover from the
sound-wave behaviour $A(\theta)=-\cos\theta$ to the single-particle
case  $A(\theta) = 1$ as the excitation
wave vector $k$ explores the
Bogoliubov dispersion relation  
$\epsilon_k =  
\sqrt{\epsilon_k^0(\epsilon_k^0+2\mu)}
$. 
In the following, we will derive the corresponding analytical expressions for the
relevant envelope function 
$A(k\xi,\theta)$ and the position of the scattering node,
$\theta_0(k\xi)$. 
}

\section{Variational theory}
Since the scattering node is clearly present in the hydrodynamic regime, we choose the 
density-phase representation $\Psi(\r,t) = \sqrt{n(\r,t)} \exp\{i
\phi(\r,t)\}$ and start with the
grand canonical energy functional 
\cite{Giorgini1994}
\begin{align}\label{Efull.eq}
E [n,\phi] = \int d^dr \biggl\lbrace  & 
\frac{\hbar^2}{2m}\left[ \bigl(\vnab \sqrt{n} \bigr)^2 + n(\vnab
\phi)^2\right]  \\ \nonumber &  
+ (V(\r)- \mu) n
+\frac{g}{2}n^2	\biggr\rbrace \ . 
\end{align}
\revision{The chemical potential $\mu$ determines the total number of
particles and introduces the healing length $\xi=\hbar/\sqrt{2m\mu}$ as the length
scale on which the condensate can respond to a spatial perturbation. 
The interaction constant $g>0$ stabilises the superfluid
behaviour of the condensate. }
The external potential $V(\r)$ shall describe the local impurity, with the influence of the very shallow trap in the centre of the BEC being negligible. 
In order to describe the dynamics of Bogoliubov
excitations in the presence of an impurity potential, we use a four-step
procedure (i-iv), equivalent in spirit to \cite{Giorgini1994}, but
with results somewhat more useful in the present context.

(i) The condensate 
ground state density $n_0(\r)$ and phase $\phi_0$ are determined in
the presence of the external potential $V(\r)$ as the saddle-point
solution of the mean-field energy functional \eqref{Efull.eq}, 
\begin{equation} \label{saddle.eq}
0 = \left.
\fder{E}{n}\right |_{n_0,\phi_0}, \qquad 0=\left.\fder{E}{\phi}\right|_{n_0,\phi_0}. 
\end{equation}
One finds that the kinetic
energy is always minimised by a spatially homogeneous
phase $ \vnab \phi_0=0$, i.e., absence of
superfluid flow. 
The ground-state density $n_0(\r)$ as function of $V(\r)$ solves the
stationary equation 
\begin{equation} \label{n0.eq} 
-\frac{\hbar^2}{2m} \frac{(\nabla^2\sqrt{n_0})}{\sqrt{n_0}} + g n_0 =
\mu - V .
\end{equation}

(ii) Density fluctuations $\delta n(\r, t)= n(\r,t)-n_0(\r)$ and phase
fluctuations $\delta \phi(\r, t)= \phi(\r, t) - \phi_0$ are 
conjugate variables that obey the coupled equations of motion 
\begin{equation} 
\label{eqGPE}
-\hbar \pder{\delta \phi}{t} = \fder{F}{(\delta n)}, \qquad  
\hbar \pder{\delta n}{t}  = \fder{F}{(\delta \phi)}. 
\end{equation}
The relevant energy functional $F$ is 
obtained by a quadratic expansion
$E=E_0+F[\delta n,\delta\phi]$ around the
saddle point and reads 
% \begin{widetext}
% \begin{equation} 
% \label{Ffull.eq}
% F [\delta n,\delta \phi] = \frac{1}{2} \int d^dr \biggl\lbrace  
% \frac{\hbar^2}{4m}\left[ \left(\vnab \frac{\delta n}{\sqrt{n_0}} \right)^2
% + \frac{(\nabla^2\sqrt{n_0})}{n_0^{3/2}} \delta n ^2 + 4 n_0 (\vnab
% \delta \phi)^2\right]  
%  + g \,\delta n^2	\biggr\rbrace \ . 
% \end{equation}
% \end{widetext}
% \begin{floatequation}\addtocounter{equation}{-1}
% \mbox{\textit{see eq.~\eqref{Ffull.eq}}}\nonumber
% \end{floatequation}
\begin{align} 
\label{Ffull.eq}
F [\delta n,\delta \phi] = \frac{1}{2} {\int} d^dr \biggl\lbrace  
\frac{\hbar^2}{4m}&\biggl[ \left(\vnab \frac{\delta n}{\sqrt{n_0}} \right)^2
+ \frac{(\nabla^2\sqrt{n_0})}{n_0^{3/2}} \delta n ^2 \nonumber \\
&+ 4 n_0 (\vnab
\delta \phi)^2\biggr] + g \,\delta n^2 \biggr\rbrace  .
\end{align}
In this formulation, the external impurity
potential $V(\r)$ affects the fluctuations only through its imprint
(via \eqref{n0.eq}) on the
ground-state density $n_0(\r)$, and this visibly in a
highly nonlinear manner. 

(iii) 
Since we wish to calculate the scattering amplitude
to linear order in $V$,  it also suffices to know the ground-state density to the same 
order. As shown in \cite{sanchez-palencia06}, by linearising \eqref{n0.eq} for small deviations from the
bulk density $n_\infty = \mu/g$, 
the condensate density $n_0(\r)$ can be written in
Thomas-Fermi form 
\begin{equation}\label{eqSmoothing}
n_0(\r) = n_\infty [1 - \tilde v(\r)].
\end{equation}
Here, the smoothed dimensionless 
potential $\tilde v(\r)=\tilde V(\r)/\mu$ is a convolution of the bare
potential $V(\r)$ by a Green's function with a very simple form
in $k$-space: 
\begin{equation}
\tilde v_{\k} = 
\frac{V_{\k}/\mu}{1+k^2 \xi^2 /2}. 
\end{equation}
This formula, derived in different notations already some time ago (eq.~(11) in
\cite{Giorgini1994}), shows that Fourier components of the effective
potential with $k\gg 1/\xi$ are
suppressed. In other words, the condensate does not follow
features of the potential varying on a length scale shorter than the
healing length $\xi$. 
%\revision{Considering the condensate in the centre of the trap as
%homogeneous, the potential $V$ will be given by the impurity in the
%following.}

Using the smooth ground-state density \eqref{eqSmoothing} in
the energy functional \eqref{Ffull.eq} 
and developing all terms to linear order in
$\tilde v(\r)$, 
we can write $F = F^{(0)} +
F^{(1)}$  as the sum of two terms: the energy of excitations of the
homogeneous bulk condensate with density $n_\infty=\mu/g$, 
\begin{equation} \label{FO.eq}
F^{(0)}%[\delta n,\delta\phi]
= \int d^dr \left\{ % 
\frac{\hbar^2}{2m} \left[ \frac{(\vnab \delta n)^2}{4n_\infty}  +
n_\infty (\vnab \delta \phi)^2 \right] + \frac{g}{2} \, \delta n^2\right\}, 
\end{equation}
and a perturbation where the smoothed 
impurity potential couples linearly to several gradient terms: 
\begin{equation} 
F^{(1)} \label{F1dphi.eq}%[\delta n,\delta\phi] 
= \int d^dr \, \tilde v(\r) \frac{\hbar^2}{2m}
\left[
\frac{(\vnab \delta n)^2-\nabla^2\delta n^2}{4n_\infty} \revision{-} n_\infty(\vnab \delta \phi)^2 
\right]. 
\end{equation}
\revision{This scattering term $F^{(1)}$ is quadratic in the
fluctuations and linear in the external potential and thus goes beyond
Huang and Meng's theory \cite{Huang1992}.}

(iv) The free-space contribution $F^{(0)}$ can be diagonalised by going into Fourier
modes followed by the Bogoliubov transformation 
\begin{align}\label{eqBgTrafo}
i \delta \phi_\k  &= \sqrt{\frac{a_k}{2}} (\gamma_\k - \gamma_{-\k}^*),
&
{\delta n_\k} &= \frac{\gamma_\k + \gamma_{-\k}^*}{\sqrt{2 a_k}} . 
\end{align}
Choosing $a_k  = \epsilon_k/(2 \epsilon_k^0 n_\infty)$ in terms of the
single-particle energy $\epsilon_k^0  = 
\hbar^2k^2/2m$ and 
the Bogoliubov dispersion  
$\epsilon_k %:= \hbar \omega_k 
= \sqrt{\epsilon_k^0(\epsilon_k^0+2\mu)}
%\mu \sqrt{k^2 \xi^2 \left( 2 + k^2 \xi^2\right)}
$ then indeed gives 
\begin{equation} \label{F0diag.eq} 
F^{(0)}[\gamma,\gamma^*]= \sum_\k %\int \frac{d^dk}{(2\pi)^d}  
\epsilon_k  \gamma_\k^* \gamma_\k \, .
\end{equation}
At this point, these Bogoliubov excitations could be conveniently 
quantised by imposing canonical commutation relations, 
which is not needed for the present purpose such that we continue to treat the
$\gamma_\k^{(*)}$ as complex field amplitudes.  

\begin{figure}%[bthp]
\begin{center}
\mbox{(a)\hspace{-3ex} \includegraphics[width=0.44\linewidth]{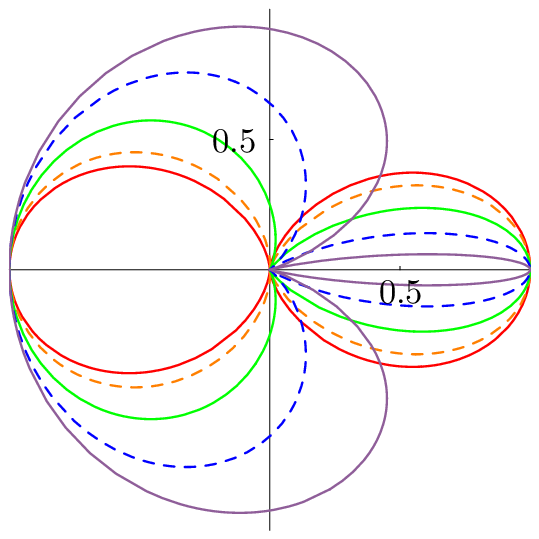}}\hfill
\mbox{(b)\hspace{-0.3ex} \includegraphics[width=0.44\linewidth]{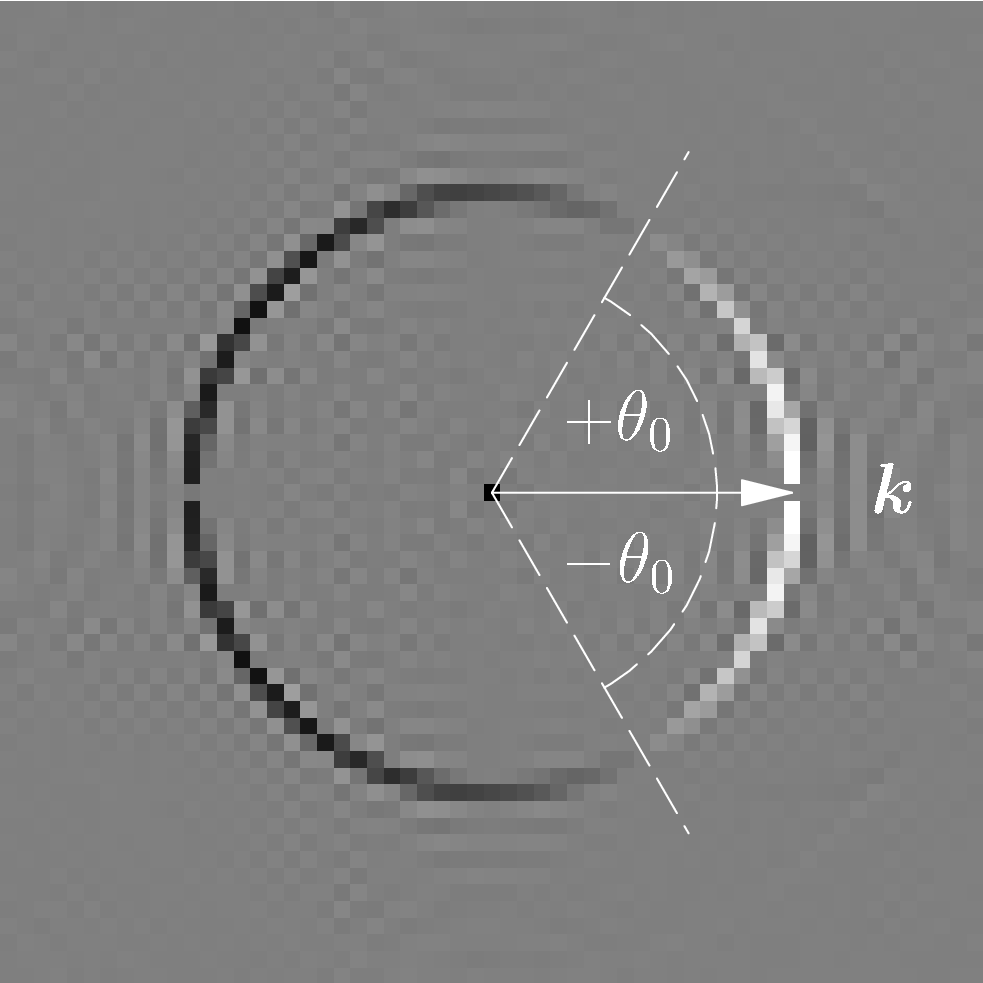}}
\caption{
(a) Polar plot  $[A(k\xi,\theta)]^2$ of the angular envelope function 
\eqref{A.eq} for  $k\xi= 0.2 \text{ (red)}, 0.5, 1,2,5 \text{
(violet)}$. %\revision{Maybe we should drop the dashed lines and add a plot with a really large $k\xi$, e.g. $k\xi=20$. This should make the isotropic case more obvious.} 
The envelope is close to a dipole radiation (p-wave) pattern
for sound waves $k\xi\ll1$, and
tends to an isotropic (s-wave) pattern for single-particle excitations
$k\xi \gg 1$. In
the intermediate regime, backscattering is favoured over 
forward scattering. 
(b) Density plot of $\Im[\gamma_{\k'}]$, the imaginary part of 
the Bogoliubov-transformed amplitude $\gamma_{\k'} = u_{k'}
\delta\Psi_{\k'} + v_{k'} \delta\Psi^*_{-\k'}$ from the numerical simulation of
the GP equation. 
This figure is more distinct than Fig.~\ref{figSetupDensity}(b), because
the interference of Bogoliubov waves with $\pm\k'$ is
eliminated, and clearly shows the sign change across the scattering node
$\theta_0$. 
Numerical parameters as in Fig.~\ref{figSetupDensity}(b).
}\label{figBgAmplitude} 
\end{center}
\end{figure}

Upon Fourier-Bogoliubov transforming, the impurity scattering
contribution \eqref{F1dphi.eq} acquires the structure 
\begin{equation}\label{eqF1Bg}
F^{(1)}[\gamma,\gamma^*] = \frac{1}{L^{d/2}}
\sum_{\k,\k'} W_{\k'\k}  \gamma_{\k'}^* \gamma_\k \,,
\end{equation}
plus terms containing products $\gamma_{\k'} \gamma_\k$ and
$\gamma^*_{\k'} \gamma^*_\k$ 
which can be disregarded for scattering 
to linear order in the impurity potential $V$. 
We find that the 
elastic scattering amplitude 
as function of the on-shell momenta
$|\k|=|\k'|=k$ and the scattering angle $\theta =
\measuredangle(\k,\k')$ writes  
\begin{align}
W(k,\theta) := W_{\k'\k}\big|_{k'=k} 
= \frac{\epsilon_k^0}{\epsilon_k} \, A(k\xi, \theta) \, V(k,\theta)  \, .\label{eqScattAmpl}
\end{align}
Here, the expected potential factor
$V(k,\theta)=V_\q$ at $q=|\k'-\k|=2k\sin(\theta/2)$ is completely
factorised from a remarkably simple \emph{angular
envelope}, 
\begin{equation}\label{A.eq}
A(k\xi,\theta) = \frac{k^2\xi^2(1-\cos\theta) - \cos\theta}
                        {k^2\xi^2(1-\cos\theta) \ + \  1 \ }\, .
\end{equation}
This angular envelope, 
drawn as a polar plot for several values of $k\xi$ in
Fig.~\ref{figBgAmplitude}(a), describes the smooth transition from
sound-wave to free-particle scattering as a function of reduced
momentum $k\xi$. 
In the deep sound-wave regime $k\xi \to 0$, $A =  -\cos(\theta)$
reproduces the dipole radiation pattern predicted by the
hydrodynamic equation 
\eqref{eqMinimalWave}. 
There is always a sign change between forward and backward
scattering since $A(k\xi,\pi) = 1 =  -A(k\xi,0) $
holds independently of $k\xi$.
From \eqref{A.eq}, the resulting node of vanishing scattering
amplitude is found to be at
\begin{equation}\label{theta0.eq}
\cos \theta_0 = \frac{k^2\xi^2}{1+k^2\xi^2} \,.
\end{equation}
In the particle regime $k\xi\gg1 $, the nodes
$\pm\theta_0$ shift to the forward direction  
such that $A(k\xi,\theta)$ converges pointwisely to the isotropic
single-particle envelope $A=1$ as $k\xi\to\infty$.  
% except for the exact forward-scattering
%direction. 
Finally, when the healing length $\xi$ becomes larger than the system size
$L$, the node angle $\theta_0 \approx \sqrt{2}/k\xi$ becomes smaller than the angular $k$-space resolution
$1/kL$. 
Then, the last contribution with negative $A(k\xi,\theta)$ is the
forward scattering element, which can be absorbed by shifting the
origin of the single-particle energy $\epsilon_k$, and we recover the
Hamiltonian for the potential scattering of free matter waves
\cite{Kuhn2005,Kuhn2007a}.

\section{Back to the simulation}
In order to confront these predictions with the 
 numerical results, 
we first of all Bogoliubov-transform the numerically calculated momentum amplitude, 
$\gamma_{\k'} = u_{k'} \delta\Psi_{\k'} \revision{+} v_{k'} \delta\Psi_{-\k'}^*$, 
with the usual coefficients 
$u_k  = (\epsilon_k + \epsilon^0_k)/(2\sqrt{\epsilon_k
\epsilon^0_k})$ and 
$v_k   = (\epsilon_k - \epsilon^0_k)/(2\sqrt{\epsilon_k
\epsilon^0_k})$. 
\revision{In the corresponding plot of $\Im[\gamma_{\k'}]$, the imaginary part of 
the Bogoliubov-transformed amplitude, shown in 
Fig.~\ref{figBgAmplitude}(b), 
one can clearly see the amplitude sign
change across the scattering node.}

\begin{figure}%[bthp]
\begin{center}
\includegraphics[width=\linewidth]{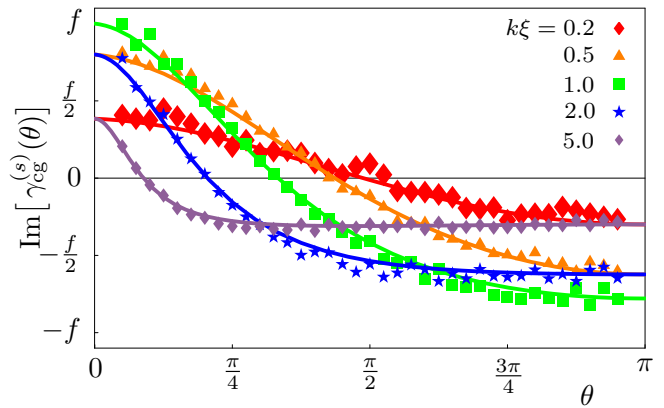}
\caption{Coarse-grained elastic scattering amplitude \eqref{eqPrediction} 
%$\Im[ \gamma^{(s)}_\text{cg}(\theta)] $ 
in units of $f = \gamma_0\, V_0/(8\xi \mu) $ for 
different values of $k\xi$, at fixed potential radius $k r_0 = 0.5$.
Symbols: Results from numerical integration of the full GP equation. Solid curves: analytical prediction \eqref{eqPrediction}.
With increasing $k$, the node moves to the left, according to \eqref{theta0.eq}. The overall amplitude has a maximum at $k\xi \approx 1$.
%The estimated errors are indicated by the size of the symbols. 
}\label{figThetaPlot} 
\end{center}
\end{figure}

For a quantitative comparison with the numerical simulation, let now a single 
Bogoliubov excitation $\gamma_{\k}^{(0)} (t) = \gamma_0 \exp\{-i\epsilon_{k}t/\hbar\}$  with wave vector $\k$ be scattered by the impurity potential as described by the effective Hamiltonian \eqref{F0diag.eq} and \eqref{eqF1Bg}. 
In linear response (equivalent to the Born approximation of the corresponding quantum problem), 
the scattered amplitude $\gamma_{\k'}^{(s)}(t) = \gamma_{\k'}^{(s)}\exp\{-i\epsilon_{k}t/\hbar\} $ is given by  
$\gamma_{\k'}^{(s)}=  \gamma_0 G_0(\epsilon_{k},k') L^{-d/2}W_{\k'\k}$,
where  $G_0(\epsilon,k) = \left[\epsilon - \epsilon_k+i0\right]^{-1}$ 
designates the retarded Green function of the free propagation
described by \eqref{F0diag.eq}.
For a quantitative comparison with the numerical simulation in the
finite system, one can coarse-grain over a $k$-space area 
%$\left(1.5 \cdot 2\pi/L \right)^2$ 
$k \Delta \theta\Delta k$
around a point on the elastic circle at a
given angle $\theta$. The analytical prediction for the coarse-grained
imaginary part %of \eqref{gammas.eq} 
then reads
\begin{align}\label{eqPrediction}
\Im[\gamma^{(s)}_\text{cg}(\theta)] = -\pi \gamma_0 \frac{\rho_0(\epsilon_{k})}{k}  
 W(k,\theta)%\frac{\pi r_0^2 V_0}{4 \xi L \mu} \,
%\frac{- k \xi}{1 + k^2 \xi^2}\,  A(k\xi, \theta)\, e^{-k^2 r_0^2 \sin^2(\theta/2)} \, .
\end{align}
in terms of the density of states per unit area, 
$\rho_0(\epsilon)= -1/(\pi L^{2})\sum_\k \Im[G_0(\epsilon,k)]$.
Coarse-graining similarly the numerical results on the elastic circle, 
we can plot together
both amplitudes as function of the scattering angle $\theta$ at
various wave vectors $k\xi$, see Fig.~\ref{figThetaPlot}. The agreement
is clearly very good, with residual numerical 
scatter around the analytical curves due to transients and boundary effects.
%partial waves backscattered from the system's boundaries.

The overall magnitude of the scattering amplitude in
Fig.~\ref{figThetaPlot} first grows and then decreases as $k\xi$
crosses from the sound-wave to the particle-regime, with 
most efficient scattering for wave vectors $k\xi\approx 1$. This
behaviour results from two competing scalings: the Bogoliubov
scattering amplitude $W(k,\theta)\propto \epsilon_{k}^0/\epsilon_{k}$ is
proportional to $k$ for $k\xi\ll 1$ and saturates to a constant for $k\xi \gg1$. The factor $\rho_0(\epsilon_{k})/k$ on the other hand
is proportional to the inverse group velocity $(\partial k/\partial
\epsilon_{k})$ that behaves like the constant $c^{-1}$ for sound waves
$k\xi\ll 1 $ and decreases as 
$k^{-1}$ for particles  $k\xi \gg1$. The product of both contributions in
\eqref{eqPrediction} therefore has limiting behaviour $k$ and
$k^{-1}$, respectively, 
with a scattering maximum around the crossover $k\xi \approx 1$ from phonons to particles.

Note that Fig.~\ref{figBgAmplitude}(b) 
is much clearer than
Fig.~\ref{figSetupDensity}(b), because there Bogoliubov
excitations with opposite $\pm \k'$ interfere in 
the wave function densities $|\delta\Psi_{\k'}|^2$.
One may wonder why the superposition of nodes stemming from opposite wave vectors  
still gives a density dip as clear as in  
Fig.~\ref{figSetupDensity}(b). In fact, in the single-particle case $k\xi\gg 1$ the ratio
$v_k/u_k$ tends to zero such that only the node of one component is observed, whereas  
for sound waves $k\xi\ll 1$ both components contribute equally, but 
now with symmetric nodes at $\pm \frac{\pi}{2}$ that
superpose exactly. This node robustness should facilitate the
experimental observation.

\section{Experimental realisability}
We propose our theoretical predictions to be experimentally tested following the numerical setup: A moderately strong, blue-detuned laser is focused 
perpendicularly through a 2D condensate \cite{Gorlitz2001,Hadzibabic2006} without depleting the condensate entirely. 
Then Bogoliubov excitations are imprinted optically 
\cite{Stamper-Kurn1999,Vogels2002,Steinhauer2002,Steinhauer2003} and observed in a subsequent time-of-flight measurement at time $t_0$.
If necessary, the sensitivity to certain $\k$-components can be greatly improved using Bragg spectroscopy \cite{Brunello2000,Vogels2002}.
In both cases, the total momentum
distribution  $|\Psi_\k(t_0)|^2=|\Psi^0_\k + \delta\Psi_\k(t_0)|^2$ at time $t_0$ is accessible. This is an oscillating quantity since the inverse Bogoliubov
transformation 
$\delta\Psi_\k(t) = u_k \gamma_\k(t) - v_k \gamma_{-\k}^*(t)$
superposes components of Bogoliubov waves scattered into opposite
directions with conjugate phases. 
The excitations $\delta\Psi_\k(t)$ live on the stationary background
of the impurity-deformed  
condensate ground state $\Psi_\k^0 \propto %-
\tilde v_\k$
given by
\eqref{eqSmoothing} such that the time-of-flight density would read  
$|\Psi_\k(t_0)|^2=(\Psi^0_\k)^2 + 2 \Psi^0_\k \Re\left[
\delta\Psi_\k(t_0)\right]  + |\delta \Psi_\k|^2$. 
If a time average is performed, the linear oscillating term drops out, and 
the density $|\delta \Psi_\k|^2$ can be extracted by subtracting the
ground-state density. 
Choosing an appropriate measurement time $t_0$ can reveal the amplitude $\Re\left[
\delta\Psi_\k(t_0)\right] + {\mathcal O}(\delta\Psi^2)$ on the smooth
background $\Psi^0_\k$ with a better signal-to-noise ratio.
The experiment could notably test the limits of validity of the weak-scattering linearisation and more generally the breakdown of mean-field behaviour.  

\section{Conclusions}

A variational treatment has allowed us to derive a simple impurity-scattering
Hamiltonian that governs the dynamics of \revision{Bogoliubov
excitations in Bose superfluids in presence of a weak external
potential}.  
Remarkably, the single-scattering amplitude
factorises into the impurity part and an angular envelope that describes
the continuous transition from wave to particle behaviour as function of excitation momentum.
\revision{Due to this factorisation, the theory is independent of the
actual shape of the potential and can also be employed to describe
disordered systems. In particular, we plan to generalise recent results on the disorder-induced localisation
of Bogoliubov excitations in 1D \cite{Bilas2006,Lugan2007a} to higher
dimensions. 
Especially the 2D case promises to be interesting, because this is the
lower critical dimension for the Anderson model of noninteracting
particles in a random potential \cite{Abrahams1979}.} 
%The implications of this anisotropic scattering with preferred
%backscattering for disorder-induced localisation in 2D remain to be
%investigated.

\acknowledgements
Financial support from DFG, BFHZ-CCUFB, and DAAD is gratefully
acknowledged. We thank V. Gurarie for explaining the dipole scattering
pattern with the hydrodynamic formulation that proved to be very
fruitful.

\bibliographystyle{mybst}
\bibliography{references}

\end{document}